\documentclass[12pt,a4paper,final]{iopart}
\usepackage{iopams}
\usepackage{graphicx}
\usepackage[breaklinks=true,colorlinks=true,linkcolor=blue,urlcolor=blue,citecolor=blue]{hyperref}

\begin{document}   
\title[Isotope effect in cuprates]{Isotope effect on the superconducting critical temperature of cuprates in the presence of charge order}

\author{Andr\'{e}s Greco$^1$, Roland Zeyher$^2$}
\address{$^1$Departamento de F\'{\i}sica, Facultad de Ciencas Exactas
e Ingenier\'{\i}a and IFIR(UNR-CONICET), Boulevard 27 de Febrero 210 bis, 2000 Rosario, Argentina}

\address{$^2$Max-Planck-Institut f\"ur Festk\"orperforschung,
             Heisenbergstrasse 1, D-70569 Stuttgart, Germany}

\eads{\mailto{agreco@fceia.unr.edu.ar},\mailto{R.Zeyher@fkf.mpg.de}}

\date{\today}

\begin{abstract}
Using the large-$N$ limit of the $t$-$J$ model and allowing also for phonons and the electron-phonon
interaction we study the isotope effect $\alpha$ for coupling constants appropriate
for YBCO. We find that $\alpha$ has a minimum at optimal doping and increases strongly (slightly)
towards the underdoped (overdoped) region. Using values for the electron phonon interaction from the local
density approximation we get good agreement for $\alpha$ as a function of $T_c$ and doping $\delta$
with recent experimental data in YBCO. Our results strongly suggest that the large increase 
of $\alpha$ in the underdoped region is (a) caused by the shift of electronic spectral density from low to 
high energies associated with a competing phase (in our case a charge density wave) and the formation of a 
gap, and (b) compatible with the small electron phonon coupling constants obtained from the local density 
approximation. We propose a similar explanation for the anomalous behavior of $\alpha$ in Sr doped 
La$_2$CuO$_4$ near the doping 1/8.  
 
 \end{abstract}


\maketitle

\section{Introduction}

The isotope effect on the superconducting transition temperature $T_c$
is one of the hallmarks of phonon-induced superconductivity in 
conventional superconductors \cite{reynolds50}. Many experiments showed that the measured isotope coefficient
$\alpha$ in these systems is near the theoretical value of 1/2 confirming the important 
role played by phonons \cite{Parks}. The isotope effect in high-$T_c$ oxides differs from
that in conventional superconductors \cite{franck94,Pringle}. Similar like $T_c$, $\alpha$ depends strongly on
doping in this case. At optimal doping, i.e., where $T_c$ assumes its largest value, $\alpha$
turns out to be very small and of the order of 0.05. Decreasing the doping $T_c$ decreases
and vanishes near the onset of long-range antiferromagnetism.
At the same time $\alpha$ increases monotonically reaching at low doping values of about 1.
Increasing the doping from its optimal value $T_c$ decreases monotonically down to zero. The
behavior of $\alpha$ in this region is presently not as clear as in the underdoped
region but seems to be constant or slightly increasing with doping \cite{franck94,Pringle}. The above characterization
of $\alpha$ applies in particular to the well-investigated Y and Bi based high-$T_c$ oxides.
The situation in Sr doped La$_2$CuO$_4$ (LSCO) is somewhat different. Large values of $\alpha$ occur near
the doping 1/8 where $T_c$ is suppressed \cite{suryadijaya05,Takagi}.

A nonzero isotope coefficient proves the involvement of phonons and the electron-phonon (EP) interaction
in the superconducting state. Since $\alpha$ assumes values near 1 in high-$T_c$ oxides, i.e., values which
are larger than in all conventional superconductors, it has been concluded \cite{franck94,bornemann91,keller05}  that phonons play 
an important role in the high-$T_c$ phenomenon. As a result theories with a strong electron-phonon coupling
and polarons have been used to explain the observed $\alpha$ \cite{kornilovitch04,macridin09,bussmann05}. On the other hand
the experiments show  that very large values of $\alpha$ occur in high-$T_c$ oxides if a competing phase with a gap or pseudogap
is present \cite{Pringle}. Theories of this kind \cite{dahm00,zeyher09} may explain $\alpha$ without assuming a strong electron-phonon coupling.
Whether the electron-phonon coupling is strong or not in cuprates is of fundamental interest.
Angle-resolved photoemission spectra show large electronic self-energies \cite{Zhou,park13} but it is not easy to decide whether
they are caused by a strong coupling to phonons \cite{Lanzara} or to spin excitations \cite{Manske}. $\alpha$, however, 
is only sensitive to phonons and not to spin excitations.  A convincing explanation of $\alpha$ thus
could also contribute to the presently controversial discussed question of the role played by phonons
in high-$T_c$ oxides.

In this paper we show that the theory of \cite{zeyher09} may explain the
recently reported doping behavior of $\alpha$ in YBCO \cite{kamiya14}. For this aim we review part of our
theory and give new expressions and discussions for $\alpha$.
In our scenario the large increase in $\alpha$ in the underdoped regime 
can be explained in the presence of a $d$ charge-density wave (CDW) state
using  small EP interaction constants as calculated in the local density approximation (LDA) \cite{heid08,Giustino,heid09}.

\section{Derivation of the expression for $\alpha$}

Our calculation of $\alpha$ is based on the Hamiltonian $H^{t-J}+H^{ep}$, where 
$H^{t-J}$ is the $t$-$J$ model and $H^{ep}$ represents the EP interaction. 
$H^{t-J}$ is given by, 
\begin{eqnarray}
H^{t-J} = -\sum_{i, j,\sigma} t_{i j}\tilde{c}^\dag_{i\sigma}
\tilde{c}_{j\sigma}  + 
 J \sum_{<i,j>} (\vec{S}_i \cdot \vec{S}_j-\frac{1}{4} n_i n_j)
+ V_c \sum_{<i, j>} n_i n_j \, .
\label{HtJ}
\end{eqnarray}
$t_{i j} = t$ $(t')$ is the hopping integral between first (second) nearest-neighbor 
sites on a square lattice;  $J$ and $V_c$ are the exchange interaction
and the Coulomb repulsion, respectively, between nearest-neighbor sites. 
$\tilde{c}^\dag_{i\sigma}$ and $\tilde{c}_{i\sigma}$ are 
creation and annihilation operators for electrons 
with spin $\sigma$ ($\sigma=\downarrow$,$\uparrow$),  respectively, excluding
double occupancies of sites. $n_i=\sum_{\sigma} \tilde{c}^\dag_{i\sigma}\tilde{c}_{i\sigma}$ 
is the electron density and $\vec{S}_i$ the spin operator. 
$<i,j>$ denotes a sum over pairs of sites $i$ and $j$. 

In the framework of the large-$N$ expansion the spin index $\sigma$ in (\ref{HtJ}) 
is extended to $N$ components, the coupling constants scaled as $t\rightarrow2t/N$, 
$t'\rightarrow2t'/N$,$J\rightarrow2J/N$ and $V_c \rightarrow2 V_c/N$, and  
the large $N$ limit is considered \cite{cappelluti99}. As a result the quasiparticle dispersion is
given by $\epsilon({\bf k}) = -2(t\delta+rJ) (\cos(k_{x})+ \cos(k_y))- 4 t'\delta \cos(k_{x}) \cos(k_y)
-\mu$ where $r= 1/N_s \sum_{\bf q}  \cos(q_x) f(\epsilon({\bf q}))$. $f$ is the Fermi function, 
$\delta$ the doping away from half-filling, $\mu$ the chemical potential, and $N_s$ the 
number of sites. 
In the following we use the lattice constant $a$ and $t$ as length and energy 
units, respectively. In addition, we take $t'/t=-0.35$ and $J/t=0.3$ which are typical values  
for cuprates.

As discussed previously \cite{cappelluti99} the above model shows instabilities with respect to a $d$ CDW \cite{Affleck,Chakravarty} 
and a superconducting phase.
The corresponding order parameters are
\begin{equation}
i \Phi({\bf k}) = -4J \gamma({\bf k}) \frac{T}{N_s}  \sum_{{\bf k'},n} \gamma({\bf k'}) g_{13} ({\bf k'},i\omega_n)
\label{OP1}
\end{equation}
\noindent and 
\begin{equation}
\Delta({\bf k}) = - 4 \tilde{J} \gamma({\bf k}) \frac{T}{N_s}\sum_{{\bf k'},n} \gamma({\bf k'})
g_{12} ({\bf k'},i\omega_n).
\label{OP2}
\end{equation}
\noindent $T$ is the temperature  and
$i\omega_n$ a fermionic Matsubara frequency.
$g_{12}$ and $g_{13}$ are the elements (1,2) and (1,3), respectively, of the $4\times4$ 
Green's function
\begin{eqnarray} 
g^{-1}(i\omega_n,{\bf k})=
\left( 
\begin{array}{c c c c}  
i\omega_n -\epsilon({\bf k}) & -\Delta({\bf k}) & -i \Phi({\bf k}) & 0  
                \\
-\Delta({\bf k}) & i\omega_n +\epsilon({\bf k}) &  0 & i \Phi({\bf {\bar k}})
                 \\
  i \Phi({\bf k}) & 0      &    i\omega_n -\epsilon({\bf {\bar{k}}})  &   -\Delta({\bf {\bar k}})
                  \\
 0  &    -i \Phi({\bf {\bar k}}) &  -\Delta({\bf {\bar k}}) &  i\omega_n +\epsilon({\bf {\bar k}})    
 \end{array} \right)
\label{matrix}
\end{eqnarray} 
\vspace{0.5cm}

\noindent with the abbreviation ${\bf {\bar k}} = {\bf k} - {\bf Q}$
where ${\bf Q} = (\pi,\pi)$ is the wave vector of the CDW.  
In (\ref{OP1}) and (\ref{OP2}) we have used the fact that the most stable solutions for $\Delta({\bf k})$ and 
$\Phi({\bf k})$ have $d$-wave symmetry, i.e., $\Delta({\bf k})=\Delta \gamma({\bf k})$ and 
$\Phi({\bf k}) = \Phi \gamma({\bf k})$
with $\gamma({\bf k})=(\cos(k_{x})-\cos(k_y))/2$.  In (\ref{OP2}) $\tilde{J}=J-V_c$ where we have introduced a 
Coulomb repulsion $V_c=0.2J$ between nearest-neighbor sites to prevent 
an instability of the CDW phase towards phase separation at low doping.

\begin{figure}
\begin{center}
\includegraphics[width=7.cm,angle=270]{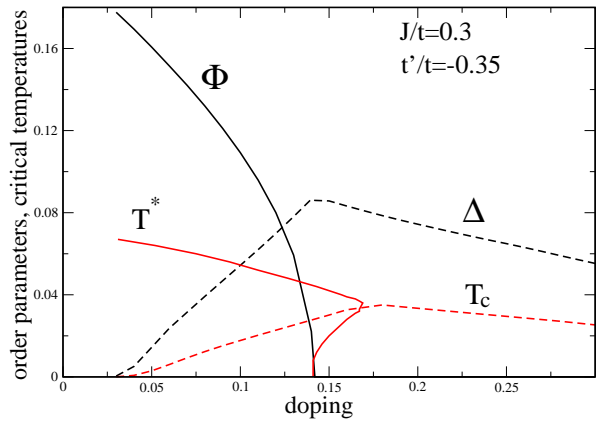}
\end{center}
\caption{Zero-temperature order parameters $\Phi$ and $\Delta$ and the critical temperatures $T^*$
and $T_c$ as a function of doping.
}
\end{figure}
For convenience we have reproduced in figure 1 previous results \cite{zeyher09} 
for $\Phi$ and $\Delta$ at $T=0$, and for
$T_c$ and $T^*$, as a function of doping. $T^*$ is  
the temperature where the CDW phase develops. The phase diagram 
is qualitatively similar to the experiments, i.e., there is
a dome-like behavior for $T_c$ with a
maximum value around
$\delta \sim 0.16$ where the CDW appears. 
Similar as in the experiments \cite{hufner08,norman05} $\Phi$ and $\Delta$ compete and coexist with each other  
at low temperatures. 
Using $t=400$ meV the resulting values for 
$T_c$ and $T^*$ compare well with the experimental ones.

Next we discuss the phonon-induced interaction $H^{ep}$ between electrons which
can be written in the static limit as, 
\begin{equation}
H^{ep} = -\frac{1}{2}{
{\sum}^\prime_{{\bf k},{\bf k'},{\bf k''},{\bf k'''},\sigma,\sigma'}}
V({\bf k}-{\bf k'}){\tilde c}^\dagger_{{\bf k}\sigma}{\tilde c}_{{\bf k'}\sigma}
{\tilde c}^\dagger_{{\bf k''}\sigma'} {\tilde c}_{{\bf k'''}\sigma'}.
\label{V}
\end{equation}
The prime at the summation sign means that
${\bf k}-{\bf k'}+{\bf k''}-{\bf k'''}$ must be equal to a reciprocal
lattice vector. 
In the following the $d$-wave part of $V({\bf k}-{\bf k'})$ will be important. It
is obtained by replacing $V({\bf k}-{\bf k'})$
by $4V \gamma({\bf k})\gamma({\bf k'})$ which defines the $d$-wave 
coupling constant $V$ for a phonon-induced nearest neighbor interaction.

An expression for $\alpha$ has been given in \cite{zeyher09}. There it has also been shown that two 
simplifications can be made without changing much the results. First, one may neglect the influence of phonons on $T^*$.  
Secondly, the EP interaction yields a contribution to the pairing but also one to
the quasi-particle weight Z. If the general question is studied whether phonons increase or decrease $T_c$ both effects
are present and compete with each other. Numerical calculations indicate that generically the second effect dominates so
that $T_c$ decreases \cite{Macridin}.  Our aim, however, is not to determine the change in $T_c$ when the EP interaction is turned on
but when the ionic mass $M$ is changed.
Writing $Z=1+\lambda_s$ it is well known that the dimensionless EP coupling constant in
the $s$-wave channel, $\lambda_s$, is independent of $M$.  The same is then true also for $Z$. Since there is good
evidence that the EP coupling in cuprates is rather small \cite{heid09} 
we may even use in the following the approximation $Z=1$. 

Introducing a phonon cutoff $\omega_0$ and the cutoff function
$\Theta_n(\omega_0-|\omega_n|)$, the gap equation reads
\begin{eqnarray} 
\Delta({\bf k},i\omega_n)&=&-4 \tilde{J} \gamma({\bf k}) \frac{T}{N_s} \sum_{{\bf k'}, n'} \gamma({\bf k'}) 
g_{12}({\bf k'},i \omega_{n'}) \nonumber \\
&&-4 V \gamma({\bf k}) \Theta_n \frac{T}{N_s} \sum_{{\bf k'}, n'} \Theta_{n'} \gamma({\bf k'}) g_{12}({\bf k'},i \omega_{n'}).
\label{delta1}
\end{eqnarray} 

\noindent The condition for $T_c$ 
can be written as,

\begin{equation}
(1+F_{11})(1+F_{22})-F_{12}^2=0
\label{FF}
\end{equation}

\noindent where 
\begin{equation}
F_{11}=-2\tilde{J} \int_0^\infty d\omega \frac{N_d(\omega)}{\omega}
\tanh(\frac{\omega}{2T_c}),
\end{equation}

\begin{eqnarray} 
F_{12}&=&-2 \sqrt{\tilde{J}V} \int_0^\infty d\omega \frac{N_d(\omega)}{\omega} \frac{2}{\pi}  \nonumber \\
&& \cdot Im [\psi(\frac{1}{2}+\frac{i\omega}{2 \pi T_c})-\psi(\frac{\omega_0}{2 \pi T_c}+1+\frac{i\omega}{2 \pi T_c})],
\label{F12}
\end{eqnarray} 
$F_{22}=\sqrt{V/\tilde{J}} F_{12}$, and $\psi$ is the digamma function. The $d$-wave projected density 
of electronic states is given by 

\begin{eqnarray} 
 N_d(\omega) &=& \frac{1}{N_s} \sum_{\bf k} \gamma^2({\bf k}) \sum_{\alpha = 1}^2  
 \frac{E^2_\alpha (\bf k) -\epsilon^2({\bf {\bar k}}) 
-\Phi^2(\bf k)}{E^2_\alpha({\bf k})-E^2_{\bar{\alpha}}({\bf k})} \nonumber \\ 
&& \cdot (\delta(\omega -E_\alpha({\bf k}))+\delta(\omega+E_\alpha({\bf k}))),
\label{Nd}
\end{eqnarray} 
with $\bar{\alpha} = 3-\alpha$ and      
\begin{equation}
 E_{1,2} = \epsilon_+(\bf k) \pm \sqrt{ \epsilon_-^2({\bf k}) + \Phi^2({ \bf k})}
\label{E12}
\end{equation}
 with $\epsilon_{\pm} = (\epsilon({ \bf k})\pm \epsilon({\bf {\bf {\bar {k}}}}))/2$.

\begin{figure}
\begin{center}
 \includegraphics[width=7cm,angle=270]{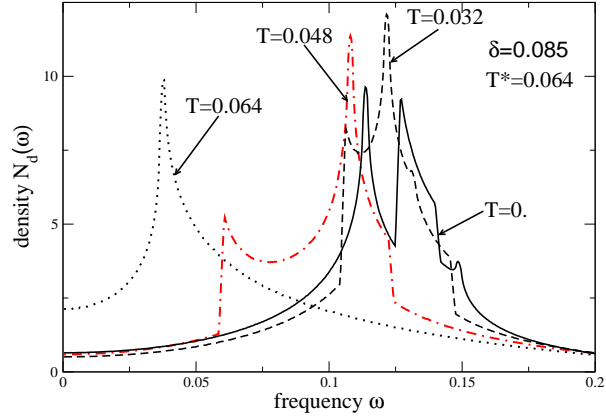}
\end{center}
\caption{
Density $N_d(\omega)$ for doping $\delta = 0.085$ and several temperatures as
a function of frequency $\omega$. 
}
\end{figure}

Figure 2 shows $N_d(\omega)$ as a function of $\omega$ for the doping
$\delta = 0.085$ and several temperatures $T$. For $T \geq T^*$ the density is dominated by a 
sharp peak at about $\omega \sim 0.04$ corresponding to the van Hove peak in the normal state. Decreasing $T$
this single peak splits into two peaks, both move towards higher energies and come closer to each other.
This behavior can be understood by noting that the main contribution in the sum over $\bf k$ comes from the surroundings
of the X-point. Then the first term under the square root in (\ref{E12}) is in general much smaller than the second
one and the square root may be expanded yielding for positive energies
\begin{equation}
E_{1,2} \rightarrow \epsilon_+({\bf k}) + \Phi({\bf k}) \pm \frac{\epsilon^2_-(\bf k)}{\Phi({\bf k})}.
\label{Eapprox}
\end{equation}
Thus one expects that $N_d(\omega)$ shows a doublet with a mean energy $\epsilon_+({\bf k}) + \Phi({\bf k})$
and a splitting energy $2\epsilon^2_-({\bf k})/\Phi({\bf k})$. With increasing $\Phi$ the splitting
decreases in agreement with the curves in figure 2.

The isotope coefficient $\alpha$ is defined by
\begin{equation}
 \alpha = \frac{\omega_0}{2T_c} \frac{\partial T_c}{\partial \omega_0}.
\end{equation}
From (\ref{FF}) follows then
\begin{equation}
 \alpha = -\frac{\omega_0}{2T_c} \frac{F'\cdot \frac{\partial F_{12}}{\partial \omega_0}}
 {\frac{\partial F_{11}}{\partial T_c}(1+F_{22}) +F'\cdot \frac{\partial F_{12}}{\partial T_c}},
\end{equation}
with
\begin{equation}
 F' = (1+F_{11})\sqrt{V/\tilde{J}}-2F_{12}.
 \label{F'}
\end{equation}
For a weak EP coupling constant $V$ $\alpha$ reduces to
\begin{equation}
 \alpha = \frac{\omega_0}{T_c} \frac{F_{12} \partial F_{12}/\partial \omega_0}{\partial F_{11}/\partial T_c},
\label{alpheinfach} 
\end{equation}
in agreement with (18) of \cite{zeyher09}.  

\begin{figure}
\begin{center}
\includegraphics[width=7.cm,angle=270]{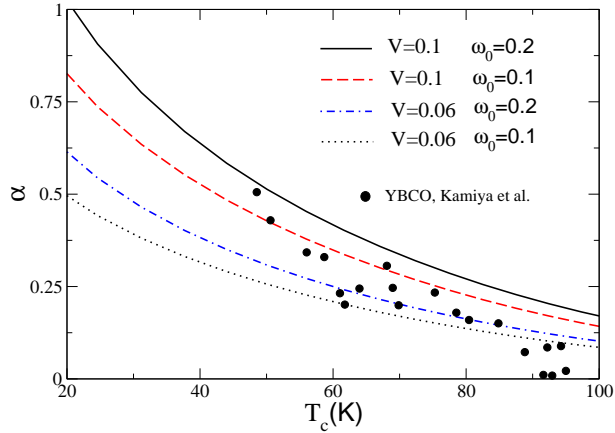} 
\end{center}
\caption{
Isotope coefficient $\alpha$ as a function of $T_c$ for different EP couplings $V$
and phonon frequencies  $\omega_0$. The filled circles are experimental points from \cite{kamiya14}
}
\end{figure}

\begin{figure}
\begin{center}
\includegraphics[width=7.cm,angle=270]{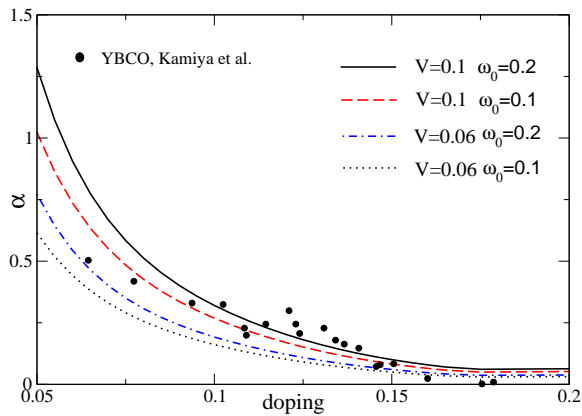} 
\end{center}
\caption{
Isotope coefficient $\alpha$ as a function of doping $\delta$ for the same parameters as in figure 3.
The filled circles are experimental points from \cite{kamiya14}
}
\end{figure}

\section{Results and discussion} 

\subsection{YBCO}

Figures 3 and 4 show $\alpha$ versus $T_c$ and $\delta$, respectively, for
$V=0.06$ and $V=0.10$, and two phonon frequencies. $\omega_0 = 0.1$ and 0.2 correspond to   
the buckling and half-breathing phonon modes in
YBCO. The solid points are experimental results from \cite{kamiya14}. They all lie in the region between the curves calculated with parameter 
values representative for cuprates. The above choice of parameter values for phonons is, of course,
unproblematic. More controversial may be the employed values for the EP coupling constants. First principles
calculation of total EP constants have been described in \cite{heid09}. The results are given in
terms of dimensionless coupling constants $\lambda_s$ and $\lambda_d$ for the $s$- and $d$-wave
channel, respectively. For each channel $\lambda$ and $V$ are related by $\lambda = V N(0)$ where 
$N(0)$ is the density of electronic states at the Fermi energy in the corresponding symmetry channel.
LDA calculations yield for YBa$_2$Cu$_3$O$_7$ $\lambda_s \sim 0.24$ and $\lambda_d \sim 0.022$ \cite{heid09}.
These values are rather small, in particular, the value for $\lambda_d$. On the other hand there is good 
evidence from several experiments that such small values are not unreasonable: Angle-resolved photoemission data
in LSCO \cite{park13,Zhao}  yielded $\lambda_s \sim 0.4$. Similarly, superconductivity-induced shifts of zone center phonons
are in good agreement with calculated LDA values \cite{Friedl,Rodri} and therefore with such small EP coupling constants. 
In figures 3 and 4 only EP coupling constants
in the $d$-wave channel enter for which we used $V=0.06$ and 0.10 which are slightly larger than the LDA values. 
They describe the experimental
points somewhat better than the bare LDA values. One should keep in mind that such adjustments
(from $\lambda_d \sim  0.02$ to $\lambda_d \sim 0.04$) should be considered as minor because we always stay 
in the region of very small EP couplings. 

It is worth remarking 
that the observed increase of $\alpha$ in the underdoped region of YBCO can be quantitatively explained
not only by employing such small values for the EP coupling constants but that a reasonable agreement between experiment
and theory requires them. As discussed above the large isotope effect found in
underdoped cuprates has been interpreted as evidence for a strong EP coupling in these systems. 
From the above analysis the conclusion is quite different: The large observed isotope shifts in
underdoped cuprates are the result of a competition of superconductivity with another ground state which produces the pseudogap.
They can be explained using the small $d$-wave EP coupling constant obtained in LDA calculations. 
Figures 3 and 4 demonstrate this for the case of a $d$ CDW state as competing state but
we expect similar results for other ground states as long as they are associated with a gap or a pseudogap.

The above calculations indicate that a pseudogap and the associated shift of the density of states
from low to high energies are responsible for the strong increase of $\alpha$ in the underdoped regime.
It is, however, clear that a reduction of $N_d(0)$ alone cannot increase $\alpha$ as long as $N_d(\omega)$
is constant on the scale of $\omega_0$. It is therefore interesting to analyze the above equations
in more detail and to find out what exactly causes the increase in $\alpha$. For the following analytical
results we will assume that the density of states $N_d(\omega)$ can be considered either as constant
or that $T_c$ is sufficiently low, i.e., we will consider the overdoped or the strongly underdoped regions.
It also will be sufficient to
consider the expression (\ref{alpheinfach}) for $\alpha$ which is valid for a weak EP coupling.

The derivative in the denominator of (\ref{alpheinfach}) may be approximated as
\begin{equation}
\frac{\partial F_{11}}{\partial T_c} = \frac{\tilde{J}}{T_c}\int_0^\infty d \epsilon \frac{N_d(2 \epsilon T_c)}
{\cosh^2(\epsilon)} \approx \frac{\tilde{J}}{T_c} N_d(0).  
\label{F11approx}
\end{equation}
The term $F_{12}$ can be written as
\begin{equation}
F_{12} = -2\sqrt{V\tilde{J}} \int_0^\infty d \epsilon N_d(\epsilon) 
[T_c \sum_n \frac{1}{\omega_n^2+\epsilon^2} -T_c \sum_{|\omega_n| >\omega_0}\frac{1}{\omega_n^2 + \epsilon^2}].
 \label{F12sum}
\end{equation}
Let us write $F_{12}$ as 
\begin{equation}
 F_{12} = \sqrt{V\tilde{J}}(F_{12}^{(1)}+ F_{12}^{(2)}),
 \label{F2}
\end{equation}
with
\begin{equation}
 F_{12}^{(1)} = \int_0
 ^{\omega_0} d \epsilon \frac{N_d(\epsilon)}{\epsilon} \tanh(\frac{\epsilon}{2T_c}), 
  \label{analytic1}
\end{equation}
\begin{equation}   
 F_{12}^{(2)} = \int_{\omega_0}^\infty d \epsilon \frac{N_d(\epsilon)}{\epsilon} \tanh(\frac{\epsilon}{2T_c})
 -\int_0^\infty \frac{N_d(\epsilon)}{\omega_0} \arctan (\frac{\omega_0}{\epsilon}).
 \label{analytic2}
\end{equation} 
In the last term in (\ref{analytic2}) the zero temperature limit has been taken.
The derivative $\frac{\partial F_{12}}{\partial \omega_0}$ acts only on the last term in $F_{12}^{(2)}$ 
 yielding
\begin{equation}
 \frac{\partial F_{12}^{(2)}}{\partial \omega_0} = - \int_0^\infty d \epsilon 
 \frac{N_d(\epsilon)}{\omega_0^2 + \epsilon^2}.
 \label{deri}
\end{equation} 
Inserting the above results into (\ref{alpheinfach}) gives 

\begin{equation} 
 \alpha = -\frac{\omega_0 V}{N_d(0)}\frac{F_{12}}{\sqrt{V\tilde{J}}}    
 \int_0^\infty d \epsilon \frac{N_d(\epsilon)}{\omega_0^2 + \epsilon^2}.
\label{alphalinear} 
\end{equation}   

Numerical evaluation of (\ref{F12sum}) shows that $F_{12}$ is practically constant
as a function of doping above optimal doping and only very slowly increasing towards lower dopings. 
Thus we may consider $F_{12}$ in (\ref{alphalinear}) as a constant. As a result we obtain
for the increase of $\alpha$ relative to its value $\alpha_0$ at optimal doping, i.e., at the onset
of the CDW,
\begin{equation}
 \alpha /\alpha_0= \frac{2}{\pi} \int_0^\infty d \epsilon \frac{\omega_0}{\omega_0^2 + \epsilon^2}
 \cdot \frac{N_d(\epsilon)}{N_d(0)}. 
 \label{alpha1}
\end{equation}
The above formula allows to understand the large increase of $\alpha$ in the underdoped regime. 
Near optimal doping $N(\epsilon)$ depends only weakly on frequency so that the density ratio
$N_d(\omega)/N_d(0)$ and therefore also $\alpha/\alpha_0$ is near 1. Below optimal doping spectral
weight is shifted from low to high frequencies which produces the pseudogap. As a result the ratio
$N_d(\omega)/N_d(0)$ is large around the pseudogap leading to large values for $\alpha/\alpha_0$.
Since $N_d(0)$ is roughly proportional to $\delta$, $\alpha/\alpha_0$ increases monotonically
with decreasing doping yielding values which may exceed by far the canonical BCS value of $1/2$.
Taking the limit $\omega_0 \rightarrow 0$ in (\ref{alpha1}) the first factor under the integral 
becomes a delta function and we obtain $\alpha/\alpha_0 = 1$. The absence of 
an enhancement of $\alpha$ for small $\omega_0$ can easily be understood: The main part
in the integral in (\ref{alpha1}) comes from the region of small $\epsilon$ well below the
pseudogap where $N_d(0)$ is small. 
The contribution from electronic spectral density shifted to
the frequency region around the gap is missing and no substantial enhancement of $\alpha$ can occur. This case
also shows that the reduction of $N_d(0)$ due to the formation of the gap does not cause an increase of $\alpha$ by itself.
Instead the shift of spectral weight from low to high frequencies near the pseudogap is responsible 
for the increase of $\alpha$. For large 
phonon frequencies $\alpha/\alpha_0$ decreases with increasing $\omega_0$. Thus one expects
that $\alpha/\alpha_0$ as a function of $\omega_0$ first increases, passes then through a maximum
and finally decreases. The curves in figure 5, calculated with the full expressions for the functions $F$,
illustrate nicely this behavior. 

\begin{figure}
\begin{center}
\includegraphics[width=7.cm,angle=270]{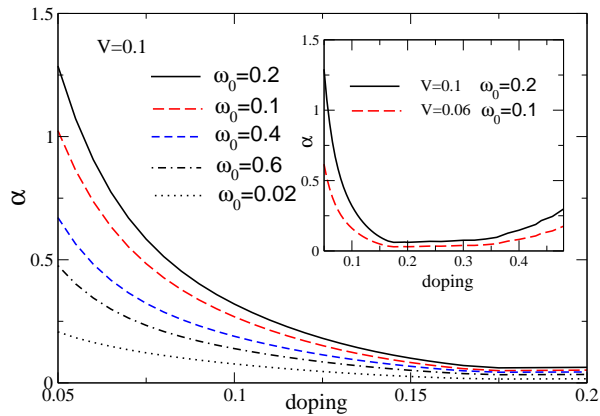} 
\end{center}
\caption{
Isotope coefficient $\alpha$ as a function of doping for $V=0.1$ and several
phonon frequencies $\omega_0$ demonstrating the nonmonotonic dependence of $\alpha$ on $\omega_0$. 
Inset: $\alpha$ over a larger doping region. 
}
\end{figure}

In the overdoped region one may assume that the density of states $N_d(\epsilon)$ is constant.
The approximations leading to (\ref{alpha1}) imply then $\alpha = \alpha_0$ throughout the overdoped
region. A better
approximation in this region is obtained by using $\omega_0$ as the cutoff in the energy integration in 
$F_{12}$. The two terms in $F_{12}^{(2)}$ cancel then exactly
and $F_{12}^{(1)}$ can be evaluated as in usual BCS theory yielding, 
\begin{equation}
 \alpha = \frac{\pi \lambda_d}{2} \;\log (\frac{1.14 \omega_0}{T_c}).
 \label{Over}
\end{equation}
A similar result has been first derived in \cite{Over}.
For optimal doping where $T_c$ is highest $\alpha$ shows a minimum with a value which for
$T_c > 1.14 \omega_0$ is weakly negative. For more realistic parameter values the logarithmic factor is
about 1 and thus $\alpha \approx \lambda_d$. The observed small values $\alpha_0 \sim 0.05$ in
cuprates with the highest $T_c$ do require similar values for $\lambda_d$ in good agreement with
the values obtained from LDA calculations. Our analytical expressions for $\alpha$ in the under- and overdoped region
used $\omega_0$ as cutoff, but one time along the imaginary and one time along the real axis. Taking the cutoff
always along the real axis is not suitable in the underdoped region because if the density peak in $N_d(\omega)$ 
coincides with the phonon frequency one obtains a spurious peak in the curve $\alpha$ versus $\delta$. Using
the cutoff along the imaginary axis we never found such an artifact and therefore used this choice of cutoff in all our
numerical calculations. 

The inset of figure 5 shows $\alpha$ versus $\delta$, calculated without approximations, over a large doping region.
In the overdoped region this function increases roughly as predicted by ({\ref{Over}). A closer look
reveals, however, that both the analytic expressions (\ref{alpha1}) for the underdoped regime 
and, to a lesser degree,  (\ref{Over}) for the overdoped regime do not well agree with the numerically
evaluated curves. One reason is that for our model $N_d({\omega})$ in Eq. (\ref{F11approx}) varies
in the energy interval $[0,2T_c]$ considerably so that the approximation proposed in that equation
is problematic. Band structure effects play thus a role in our model even at low energies producing
fluctuations in the curve $\alpha$ versus $\delta$ if calculated with (\ref{alpha1}). In the numerically exact
calculated curves in figures 3-5 such fluctuations are absent due to a properly carried out energy integration
in ({\ref{F11approx}). Though (\ref{alpha1}) and ({\ref{Over}) are thus not suitable to obtain accurate
values they nevertheless explain correctly the curves $\alpha$ versus $\delta$ at low and high dopings. 

\subsection{La and Ba doped La$_2$CuO$_4$}

Our theory can also be applied to Sr and Ba doped La$_2$CuO$_4$. La$_{2-x}$Ba$_x$CuO$_4$ 
shows a variety of phase transitions near the doping $\delta = 1/8$ \cite{Huecker}. 
$T_c$ exhibits a dip between $\delta_1 = 0.155$ and $\delta_2 = 0.095$ which 
nearly touches zero at $\delta = 0.125$ \cite{Mood,Huecker}. Between these limiting dopings
a leading phase CO with charge stripe order extends towards higher temperatures above the superconducting phase 
in a domelike manner touching $T_c$ at the end points. There are additional phases of spin stripe order
or of orthorhombic or tetragonal symmetry present which will be disregarded
in the following. Above $\delta_1$ or below $\delta_2$, $T_c$ is decreasing with increasing distance from 
these points reaching very small values near $\delta = 0.25$ and 0.05, respectively. The phase diagram
of the sister compound LSCO does not contain long-ranged phases around the doping 1/8. Nevertheless
it is probable that superconductivity competes with phases in the particle-hole channel near this
doping because $T_c$ shows a pronounced dip in this region \cite{suryadijaya05,Takagi}.
Also angle resolved
photoemission experiments find in LSCO a pseudogap which sets in at around $\delta = 0.20$ and exists and increases
in magnitude towards lower dopings. The isotope coefficient in LSCO (see figure 6) is very small at the large doping
value of 0.20 where the pseudogap forms. With decreasing $\delta$, $\alpha$ first increases slightly and
then vary rapidly reaching a maximum near the doping 1/8 with a value of about 1. Decreasing $\delta$ further
$\alpha$ decreases but settles down in the region of about 0.5.

In view of the above phase diagrams and the behavior of $\alpha$ in LSCO it is natural to assume that
the doping dependence of $T_c$ and $\alpha$ are caused by gaps or pseudogaps similar like in YBCO.
In order to transfer our results from YBCO to La$_{2-x}$Ba$_x$CuO$_4$ and LSCO
we consider the calculated $\alpha$ in YBCO as a function of
$T_c/T_{c,0}$ where $T_{c,0}$ is the optimal $T_c$ for a doping near the onset of the pseudogap.
Treating first  La$_{2-x}$Ba$_x$CuO$_4$ we can read off from figure 2 of \cite{Mood} the doping as a function of
$T_c/T_{c,0}$ where $T_{c,0}$ is the value of $T_c$ near the dopings $\delta_1$
or $\delta_2$. Identifying the two ratios for the reduction of $T_c$ one obtains $\alpha$ as a function of
doping in La$_{2-x}$Ba$_x$CuO$_4$. The same procedure can be applied to LSCO. From figure 3 in \cite{Takagi}}
one can read off $\delta$ as a function of $T_c/T_{c,0}$ where $T_{c,0}$ is the largest transition temperature
near $\delta = 0.15$. Comparing this reduction ratio with the case of YBCO one finds the doping dependence
of $\alpha$ in LSCO. Figure 6 contains the obtained curves, calculated for $V=0.1$ and $\omega_0=0.1$,
together with experimental values for LSCO from \cite{suryadijaya05}. Our calculation suggests
the following interpretation of the experimental $\alpha$ in LSCO. One has to distinguish between two pseudogaps in
LSCO. The first one is the usual pseudogap, which is observed  by angle resolved photoemission and which 
exists below $\delta \sim 0.20$ \cite{Yoshida}. This pseudogap corresponds to that occurring in underdoped cuprates and 
develops at $T^*$. At low doping superconductivity competes with the phase underlying the pseudogap
leading to the overall decrease of $T_c$ down to very low values near $\delta = 0.05$.
At the same time this decrease of $T_c$ is associated with a monotonic increase of $\alpha$ similar as in
YBCO. The second pseudogap is located between $\delta = 0.10$ and $\delta = 0.15$ and is due to static (in La$_{2-x}$Ba$_x$CuO$_4$)
or fluctuating (in LSCO) stripes. As a result $T_c$ is strongly (in La$_{2-x}$Ba$_x$CuO$_4$) or slightly
(in LSCO) suppressed. Correspondingly, the increase of $\alpha$ in figure 6 is large for the Ba and small for the Sr
doped systems. 
\begin{figure}
\begin{center}
\includegraphics[width=7.cm,angle=270]{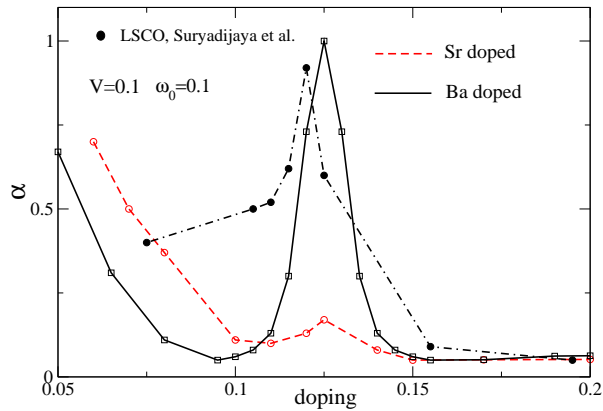} 
\end{center}
\caption{
Isotope coefficient $\alpha$  as a function of doping. Filled circles
are experimental points in LSCO from \cite{suryadijaya05}. Empty circles and squares
are calculated values for La$_{2-x}$Ba$_x$CuO$_4$ and LSCO, respectively. Corresponding points have 
been joined smoothly by lines.
}
\end{figure}
It is interesting to see that the calculated curve for LSCO shows only a small peak at doping 1/8 quite
in contrast to the experimental curve. The reason for this is that the employed experimental $T_c$ curve
exhibits only a small dip near the doping 1/8. One prediction of our calculation is that $\alpha$
in Ba doped La$_2$CuO$_4$ should show a large peak near doping 1/8. Though our calculation for $\alpha$
in LSCO is not able to get quantitative agreement with experiment we think that the shift of spectral weight
from low to higher energies associated with the formation of the pseudogap plays also a role 
in LSCO.

Finally we would like to mention that strong renormalizations
(softenings and broadenings) of phonons have been observed
in underdoped cuprates (see \cite{reznik06} and references there in).
These anomalies occur for wave vectors along the crystalline axis with
a length of about 0.3 and are seen both in acoustical \cite{letacon14}
and optical bond-stretching phonons \cite{pintscho}. They
are probably related to
the recently observed charge order in these systems \cite{letacon14}.
On the other hand extensive LDA calculations, performed for YBa$_2$Cu$_3$O$_7$,
did not yield unusual softenings \cite{heid09,reznik08} and cannot explain these
phonon anomalies. However, our main result is independent of the
validity of the LDA in cuprates and shows only
that the observed large values of $\alpha$ in underdoped cuprates
are at least compatible with the LDA values for the electron-phonon
coupling constants. The existence of the above phonon anomalies suggests 
that it is not
possible to conclude quite generally from these and other successful LDA
results \cite{Friedl,Rodri} that the electron-phonon
coupling is necessarily small in the cuprates. Thus, 
the correct explanation of the phonon anomalies
is presently an open problem and beyond the scope of our paper,
similar as their influence on T$_c$.

\section{Conclusions}

In summary, our calculations of the isotope coefficient $\alpha$ were based on a mean-field like treatment of the
$t$-$J$ model where optimal doping coincides with the onset of a charge-density wave which competes
with superconductivity in the underdoped regime and suppresses $T_c$ there. Adding phonons and the electron-phonon
coupling our model can explain the strong experimental increase of $\alpha$ in the underdoped region, using at the same time the
small electron-phonon coupling constants from the LDA. 
Thus we conclude that the large values for $\alpha$ in the underdoped regime give no evidence for a large electron-phonon
coupling in these systems but are compatible with the small LDA values once the competing phase is taken into account.
Large enhancements of $\alpha$ are found if the phonon energy $\omega_0$ and the gap $\Phi$ are comparable in magnitude and are absent for 
very small or large ratios $\omega_0/\Phi$.
Our explanation of the behavior of $\alpha$ is supported by several experimental facts: $\alpha$ assumes its minimum at optimal doping
where the competing phase sets in and strongly increases towards lower dopings in the presence of the competing phase;
in this region superconductivity forms from a state where the density of electronic states varies strongly over the scale of phonon 
energies which is quite different from 
the normal state. All these features make in our opinion explanations very improbable which are
based on a strong electron-phonon interaction, polarons or anharmonic mechanisms.   

\ack The authors thank P. Horsch and M. Bejas for a critical reading of the manuscript. R.Z. and A.G. are grateful to the
Instituto de F\'{i}sica (Rosario) and the MPI-FKF (Stuttgart), respectively, for hospitality and financial support. The authors
are in particular grateful to the late Manuel Cardona for many enlightening and stimulating discussions on high-T$_c$ superconductors. 

\section*{References}

\bibliography{biblio.bib}

\end{document}